\documentclass[12pt]{article}
\usepackage[utf8]{inputenc}
\usepackage{graphicx}
\usepackage{amsmath}

\usepackage{color}
\definecolor{cCC9900}{RGB}{204,153,0}

\usepackage{authblk}
\usepackage{epstopdf}

\title{Coherent noise source identification in multi channel analysis}
\date{\today}
\author[1]{T.\,Frisson\thanks{corresponding author: thibault.frisson@cern.ch - now at CERN}}
\author[1]{R.\,Poeschl}
\affil[1]{Laboratoire de L'acc\'elerateur Lin\'eaire (LAL), CNRS/IN2P3, Orsay, France}

\begin{document}
 \maketitle

 \section{Introduction}

The evaluation of coherent noise can provide useful information in the study of detectors. The identification of coherent noise sources is also relevant for uncertainty calculations in analyse where several channels are combined. The study of the covariance matrix give information about coherent noises. Since covariance matrix of high dimension data could be difficult to analyse, the development of analysis tools is needed. Principal Component Analysis (PCA) is a powerful tool for such analysis. It has been shown that we can use PCA to find coherent noises in ATLAS calorimeter~\cite{zitoun} or the CALICE Si-W electromagnetic calorimeter physics prototype~\cite{Adloff201197}. However, if several coherent noise sources are combined, the interpretation of the PCA may become complicated.

In this paper, we present another method based on the study of the covariance matrix to identify noise sources. This method has been developed for the study of front end ASICs dedicated to CALICE calorimeters. These calorimeters are designed and studied for experiments at the ILC~\cite{dbd}. We also study the reliability of the method with simulations. Although this method has been developped for a specific application, it can be used for any multi channel analysis.


\section{Method}

The goal is to identify and characterize dissociable noise sources in a multi-channel systems. This method cannot separated noise sources which affect exactly the same set of channels. In this case, the noises sources are processed as a single source. We consider a system with N channels. Each channel $i$ is affected by an incoherent noise source I$_{i}$ and N$_{c}$ coherent noise sources (C$^{1}_{i}$, C$^{2}_{i}$, ..., C$^{n}_{i}$). We assume that all noise source distributions are Gaussian and independant. The noise of each noise source is the standard deviation of the distribution. The Gaussian assumption will be studied in section~\ref{section:NGN}. Let $\sigma_{I_{i}}$ denote the noise of the incoherent source for channel $i$ and let $\sigma_{C^{j}_{i}}$ denote the noise of the coherent source $j$ for channel $i$. The total noise $\sigma_{i}$ in the channel i is:
\begin{equation}
\sigma_{i}^{2} = \sigma_{I_{i}}^{2} + \sum^{N_{c}}_{j=1}  \sigma_{C^{j}_{i}}^{2} \label{eq:noise}
\end{equation}

The covariance matrix element from the two channels i and k is expressed by:
\begin{equation}
  cov(i,k) = \delta_{ik}\sigma_{I_{i}}\sigma_{I_{k}} + \sum^{N_{c}}_{j=1}\sigma_{C^{j}_{i}}\sigma_{C^{j}_{k}} \label{eq:covElement}
\end{equation}
where:
\begin{equation}
  \delta_{ik} = \begin{cases}
1 & \text{ if } i = k \\
0 & \text{ if } i \neq k
\end{cases}
\end{equation}

The covariance matrix element can also be determined from the data:
\begin{equation}
  cov_{Data}(i,k) = \frac{\sum^{N_{event}}_{n=1} \left(A_{i}(n) - \mu_{A_{i}})(A_{k}(n) - \mu_{A_{k}}\right)}{N_{event}} \label{eq:covData}
\end{equation}
where $N_{event}$ is the number of events, $A_{i}(n)$ is n-th event of channel i and $\mu_{A_{i}}$ is the mean value of channel i.

We built an equation system with equation~\ref{eq:covElement} and equation~\ref{eq:covData}:
\begin{eqnarray*}
  cov(1,1) &=& cov_{Data}(1,1) \\
  &\vdots& \\
  cov(1,k) &=& cov_{Data}(1,k) \\
  &\vdots& \\
  cov(i,k) &=& cov_{Data}(i,k) \\
  &\vdots& \\
  cov(N,N) &=& cov_{Data}(N,N)
\end{eqnarray*}
The covariance matrix is symmetric thus we do not consider redundant equations. The system is composed of $\frac{1}{2} \cdot N \cdot (N + 1)$ equations and $N + N_{c} \cdot N$ parameters. The parameters are estimated by minimizing the quantity:
\begin{equation}
  \sum^{N}_{i=1}\sum^{N}_{k=i} \left(cov(i,k) - cov_{Data}(i,k)\right)^{2} \label{eq:minimization}   
\end{equation}
The minimization is realized using Minuit and the ROOT software~\cite{minuit}~\cite{root}. The minimization residual decreases as a function of N$_{c}$ and converges to a minimum. The selected N$_{c}$ is the value for which the residual reaches this minimum.


\section{Application to the SPIROC ASIC}

\subsection{The SPIROC ASIC}

The SPIROC chip is a dedicated very front-end electronics for an ILC prototype hadronic calorimeter with Silicon photomultiplier readout. SPIROC is a dual gain 36-channel ASIC which allows to measure on each channel the charge from 1 to 2000 photoelectrons and the signal timing with a 100 ps accuracy time-to-digital converter. For each channel, two low noise preamplifiers ensure the requested dynamic range with noise level of 1/10 photo-electron. The amplification ratio between the two gains is ten. An analogue memory array with a depth of 16 for each channel is used to store the time information and the charge measurement. A 12-bit Wilkinson Analogue-to-digital Converter (ADC) has been embedded to digitize the analogue memory content~\cite{spiroc}.

Pedestals are extracted using an external signal to trigger the ASIC. The low gain output and the high gain output are studied in parallel. Each acquisition contains 16 events of 72 values. In the following analysis, the high gain output of the 36 channels are the channels numbered from 0 to 35. The low gain output of the 36 channels are the channels numbered from 36 to 71.  

\subsection{Noise study}\label{section:Noisestudy}

\begin{figure}[h]
 	{\centering 
 		\includegraphics[width=0.9\textwidth]{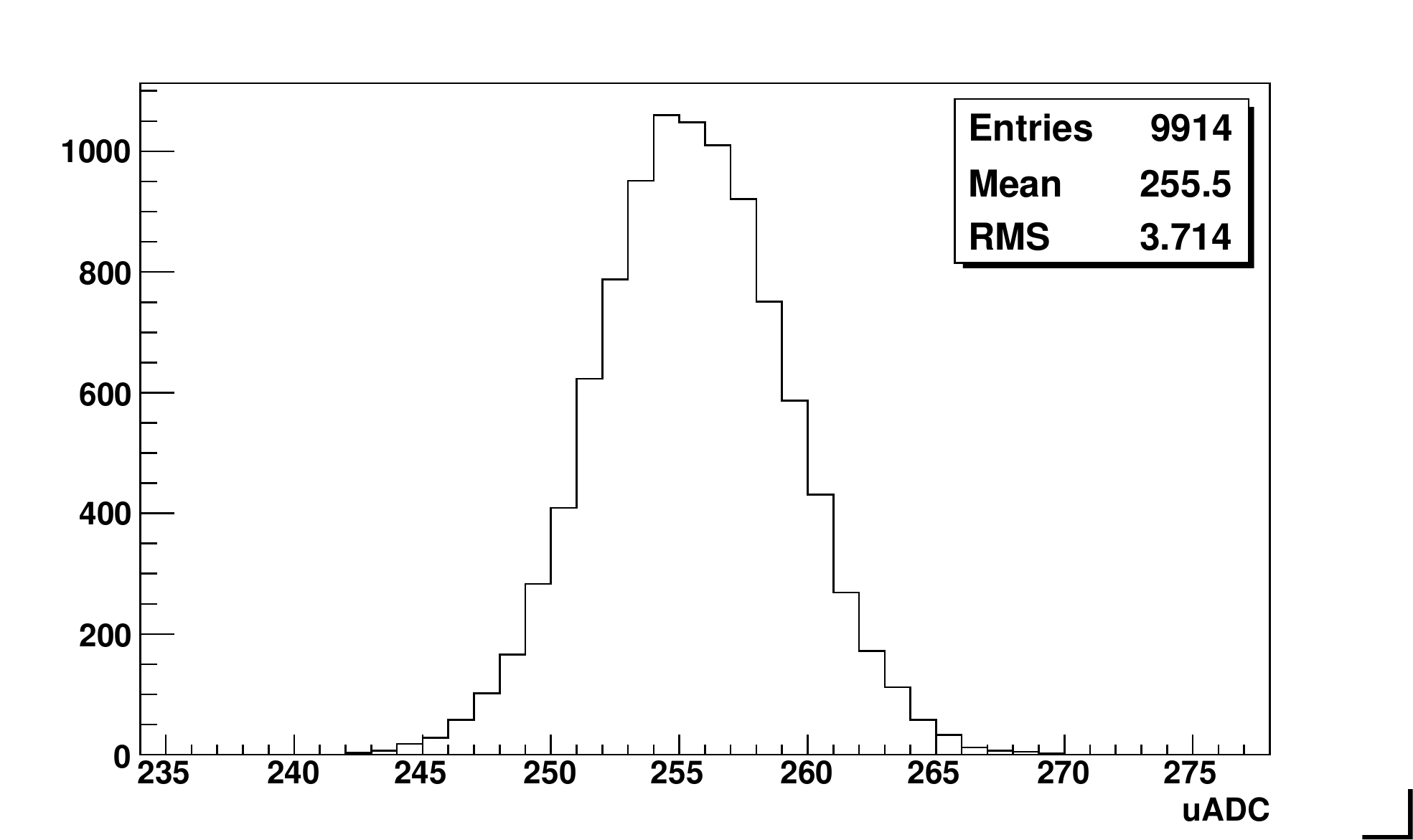}
 	\caption{Example of pedestal (channel 16)}
 	\label{fig:exPed}}
 \end{figure}

An example of measured distribution is shown in figure~\ref{fig:exPed}. The pedestal is given by the mean of the distribution, the total noise by the rms.

\begin{figure}[h]
 	{\centering 
 		\includegraphics[width=.5\textwidth]{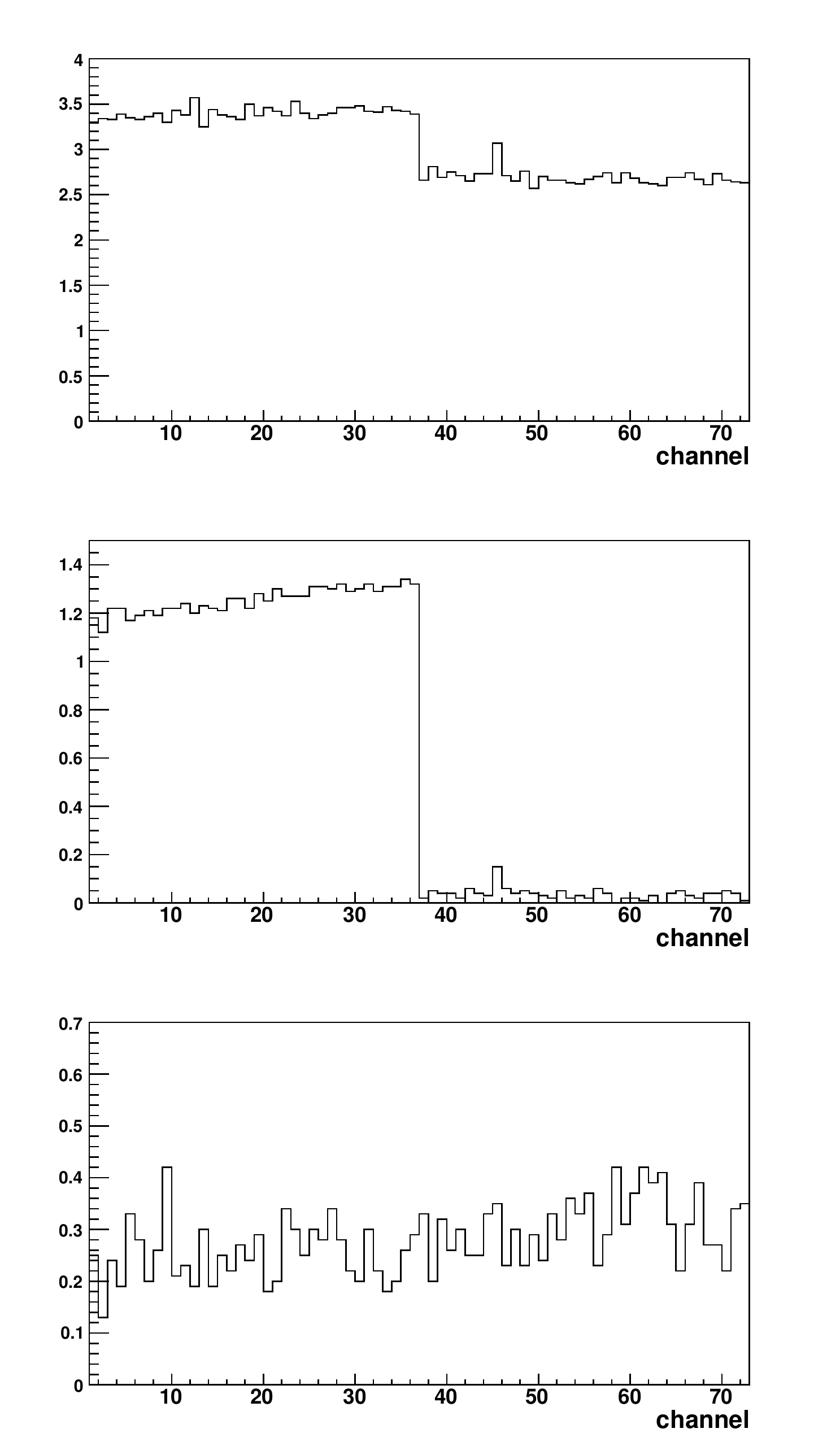}
 	\caption{Top: $\sigma_{I_{i}}$ as a function of the channel number - middle:  $\sigma_{C^{1}_{i}}$ as a function of the channel number - bottom: $\sigma_{C^{2}_{i}}$ as a function of the channel number.}
 	\label{fig:resSPIROC}}
 \end{figure}

We apply the method to a run of about 10,000 aquisitions i.e. 160,000 events. The minimization procedure converges with 2 coherent sources. The results are presented in figure~\ref{fig:resSPIROC}. $\sigma_{I_{i}}$, $\sigma_{C^{1}_{i}}$ and $\sigma_{C^{2}_{i}}$ are plotted as a function of the channel number respectively on the top, on the middle and on the bottom of the figure. The incoherent sources are found to have a mean noise $<\sigma_{I_{HG}}>$ = 3.4 uADC for high gain channels and $<\sigma_{I_{LG}}>$ = 2.7 uADC for low gain channels. The first coherent source has a mean noise $<\sigma_{C^{1}_{HG}}>$ = 1.25 uADC for high gain channels and $<\sigma_{C^{1}_{LG}}>$ = 0.04 uADC for low gain channels. This source is negligible for the low gain part of the ASIC which indicates that the noise depends on the amplification. Thus, the preamplifiers could be the noise source. We also see a slight increase of the noise as a function of the channels number. The second coherent source affects all channels. The analog signals of the two gains of all channels are converted at the same time in the ADC. Thus, the ADC could be the source of this coherent source. The mean intensity is 0.3 uADC for all channels.

\section{Simulation studies}

To study the reliability of this method, simulated data files are produced with various noise source features. We apply the method on each simulated data file and we compared the results with the source defined in the simulation.

Two criteria are chosen to estimate the robustness of the algorithm:
\begin{eqnarray}
  \Delta \sigma_{i}  &=& \sigma^{gen}_{i} - \sigma^{cal}_{i} \label{eq:critereAbs} \\
  \Delta_{rel} \sigma_{i}  &=& \frac{\sigma^{gen}_{i} - \sigma^{cal}_{i}}{\sigma^{tot}_{i}} \label{eq:critereRel}
\end{eqnarray}
where $\sigma^{gen}_{i}$ is the noise of the studied source defined in the simulation for channel $i$, $\sigma^{cal}_{i}$ is the noise of the studied source calculated with our method for channel $i$ and $\sigma^{tot}_{i}$ is the total noise of channel $i$ defined in the simulation.

100,000 events are simulated per configuration. The reference configuration is defined to have similar noise distributions than those calculated in section~\ref{section:Noisestudy}. The incoherent noise $\sigma_{I_{i}}$ of channel $i$ is defined randomly following a uniform distribution with a width 0.4 uADC and centered on $<\sigma_{I_{HG}}>$ = 3.5 uADC for the 36 high gain channels (channels from 0 to 35) and $<\sigma_{I_{LG}}>$ = 2.6 uADC for the 36 low gain channels (channels from 36 to 71). A first coherent source (C$^{1}$) is defined for the 36 high gain channels. The noise $\sigma_{C^{1}}$ = 1.2 uADC is equally distributed between the high gain channels. All channels are also affected with a second coherent source (C$^{2}$). The noise $\sigma_{C^{2}}$ = 0.5 uADC is equally distributed between the channels.

\subsection{Incoherent noises}

In this part, we study various configurations of the incoherent sources. We produce 8 configurations. $<\sigma_{I_{HG}}>$ is varied from 1.5 uADC to 8.5 uADC. $<\sigma_{I_{LG}}>$ is varied from 0.5 uADC to 7.5 uADC. The reference configuration is used for the coherent sources.

\begin{figure}[h]
 	{\centering 
 		\includegraphics[width=1.\textwidth]{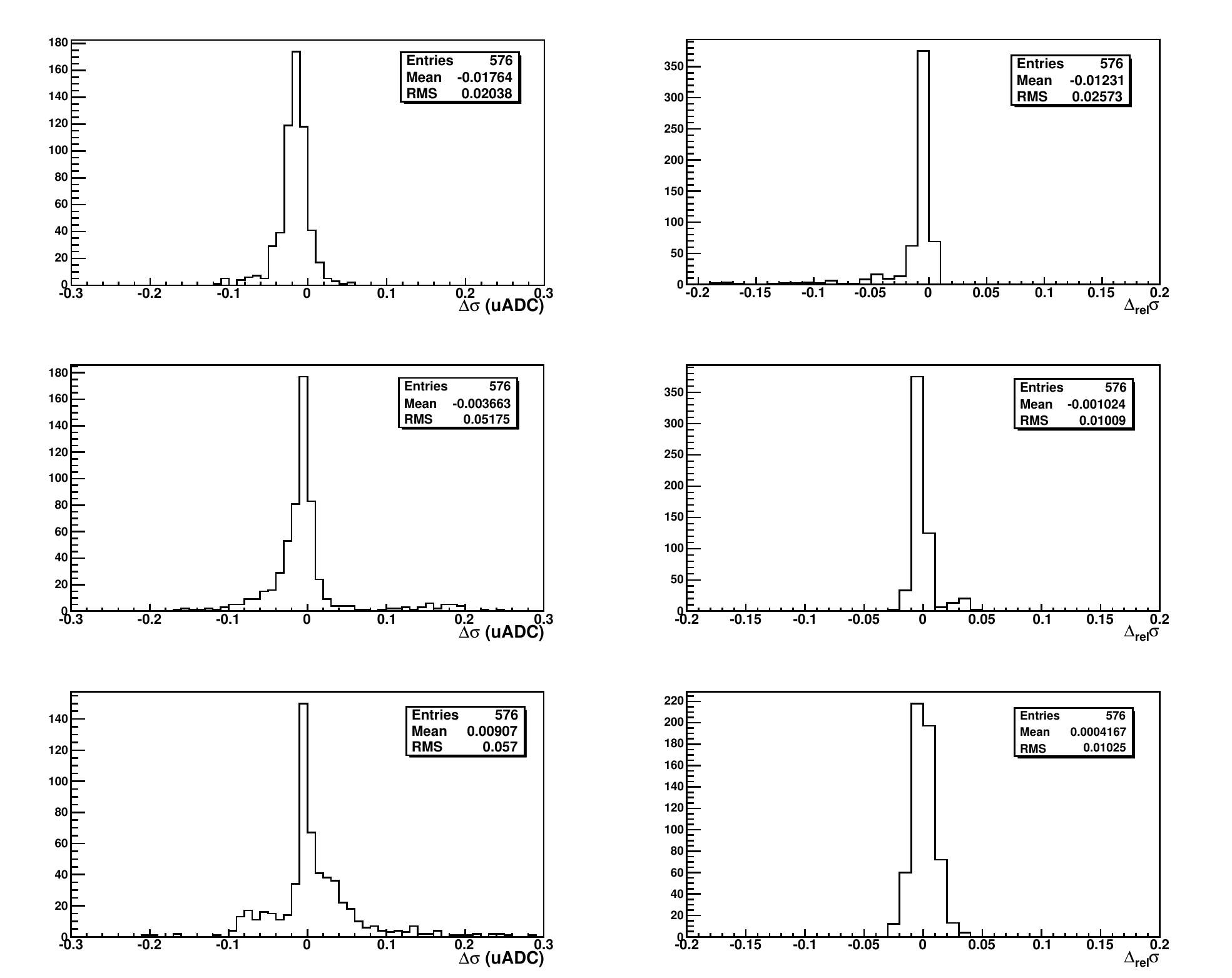}
 	\caption{Left side: $\Delta \sigma_{i}$ in ADC units - Right side: $\Delta_{rel} \sigma_{i}$ - Criteria for incoherent sources, C$^{1}$ and C$^{2}$ are plotted respectively on the top, on the middle and on the bottom.}
 	\label{fig:resInc}}
 \end{figure}

Figure~\ref{fig:resInc} shows the robustness criteria for all channels and all simulated configurations. $\Delta \sigma_{i}$ is on the left and $\Delta_{rel} \sigma_{i}$ is on the right. Criteria for incoherent sources, C$^{1}$ and C$^{2}$ are plotted respectively on the top, on the middle and on the bottom.

$\left | \Delta \sigma_{i} \right |$ is always below 0.3 uADC. The relative difference between simulated and calculated noise is always below than 5~\% exept for the incoherent source of some channels. For the configuration with $<\sigma_{I_{LG}}>$ = 0.5 uADC, $\sigma^{tot}_{i}$ is small and $\Delta \sigma_{i}$ is around 0.1 uADC. Therefore the relative difference reach about 20~\%.

\subsection{Coherent noises}

In this part, we study various configuration of coherent sources. We produced 12 configurations. $\sigma_{C^{1}_{i}}$ is varied from 0.5 uADC to 10 uADC. $\sigma_{C^{2}_{i}}$ is varied from 0.2 uADC to 10 uADC. The reference configuration is used for incoherent sources.

\begin{figure}[h]
 	{\centering 
 		\includegraphics[width=1.\textwidth]{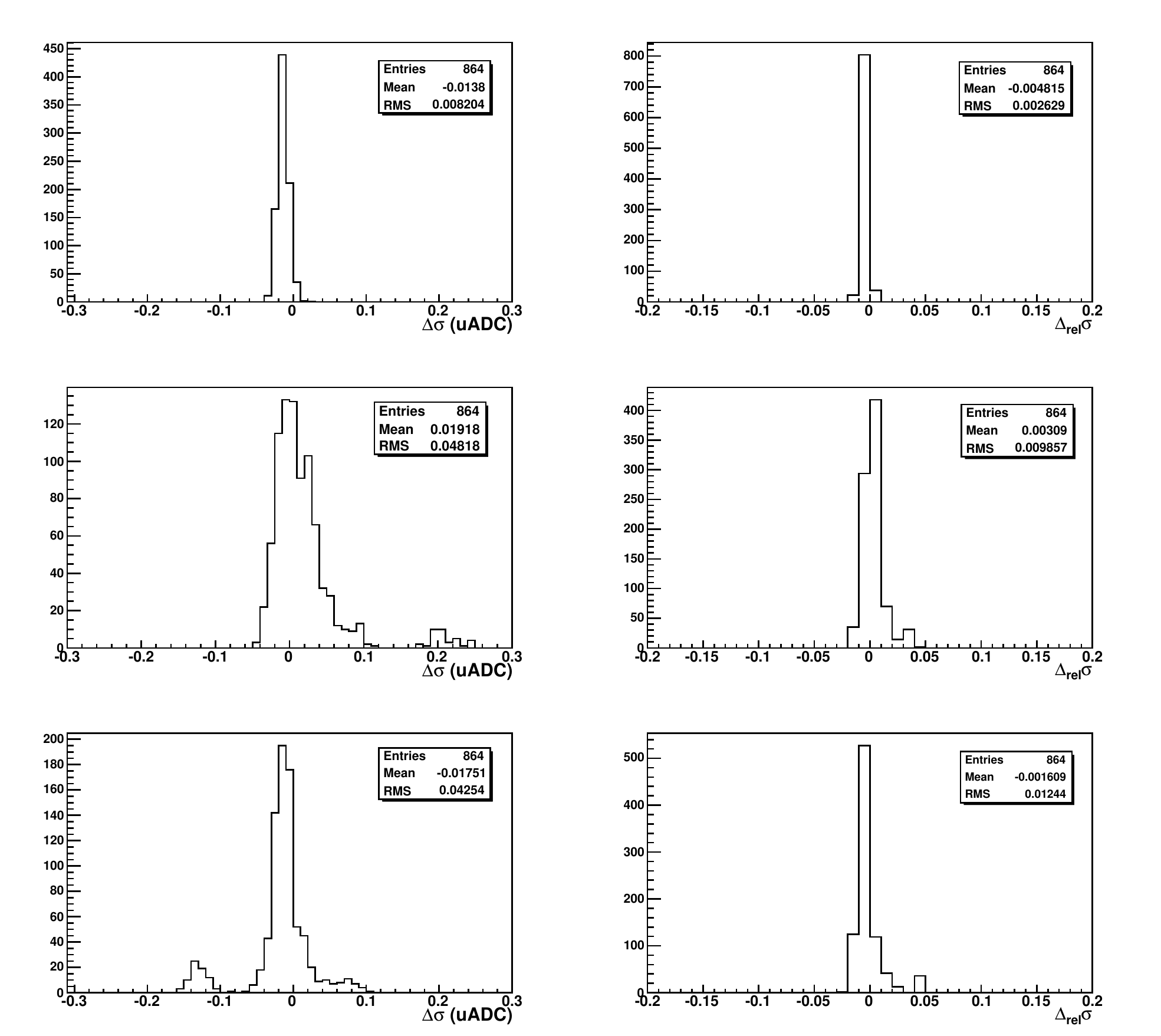}
 	\caption{Left side: $\Delta \sigma_{i}$ in ADC units - Right side: $\Delta_{rel} \sigma_{i}$ - Criteria for incoherent sources, C$^{1}$ and C$^{2}$ are plotted respectively on the top, on the middle and on the bottom.}
 	\label{fig:resCoh}}
 \end{figure}

Figure~\ref{fig:resCoh} shows the robustness criteria of the method for all channels and all simulated configurations. $\Delta \sigma_{i}$ is on the left and $\Delta_{rel} \sigma_{i}$ is on the right. Criteria for incoherent sources, C$^{1}$ and C$^{2}$ are plotted respectively on the top, on the middle and on the bottom.

\begin{figure}[h]
 	{\centering 
 		\includegraphics[width=.8\textwidth]{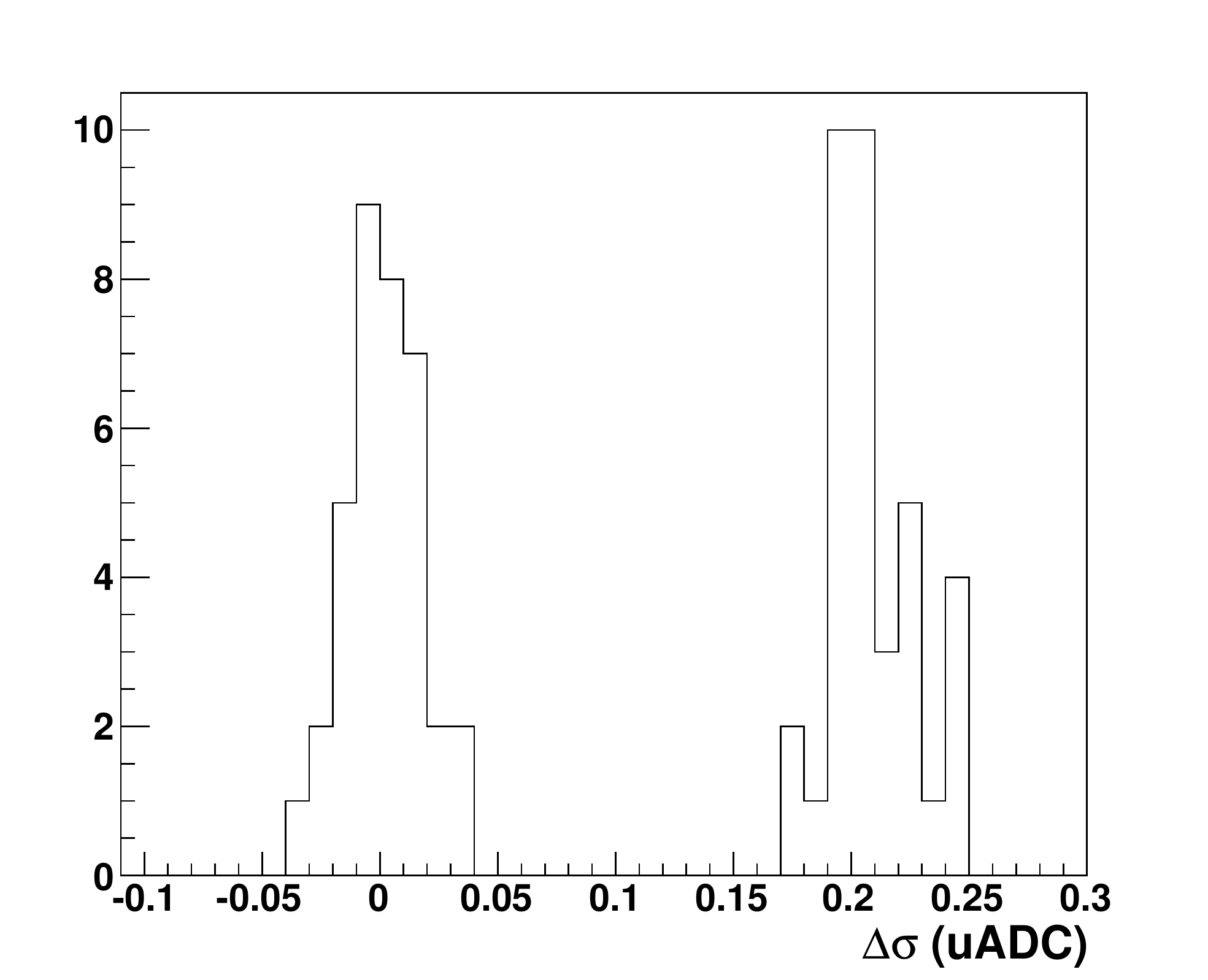}
 	\caption{$\Delta \sigma_{i}$ in ADC units for the C$^{1}$ ( $\sigma_{C^{1}_{i}}$ = 5 uADC and $\sigma_{C^{2}_{i}}$ = 0.5 uADC)}
 	\label{fig:pb}}
 \end{figure}

The relative difference between simulated and calculated noise is always below than 5~\% and $\left | \Delta \sigma_{i} \right |$ is below 0.1 uADC exept for some entries around 0.2 uADC for C$^{1}$ and around 0.12 uADC for C$^{2}$. In figure~\ref{fig:pb}, $\Delta \sigma_{i}$ of C$^{1}$ is shown for the configuration $\sigma_{C^{1}_{i}}$ = 5 uADC and $\sigma_{C^{2}_{i}}$ = 0.5 uADC. We see two peaks. The peak at 0.2 uADC corresponds to the low gain channels. The minimization reach a local minimum without ever finding the global minimum. Even if $\Delta_{rel} \sigma_{i}$ stays below 5~\% in this case, this shift could lead to a misinterpretation of the results.

\subsection{Non-Gaussian noises}\label{section:NGN}

\begin{figure}[h]
 	{\centering 
 		\includegraphics[width=.9\textwidth]{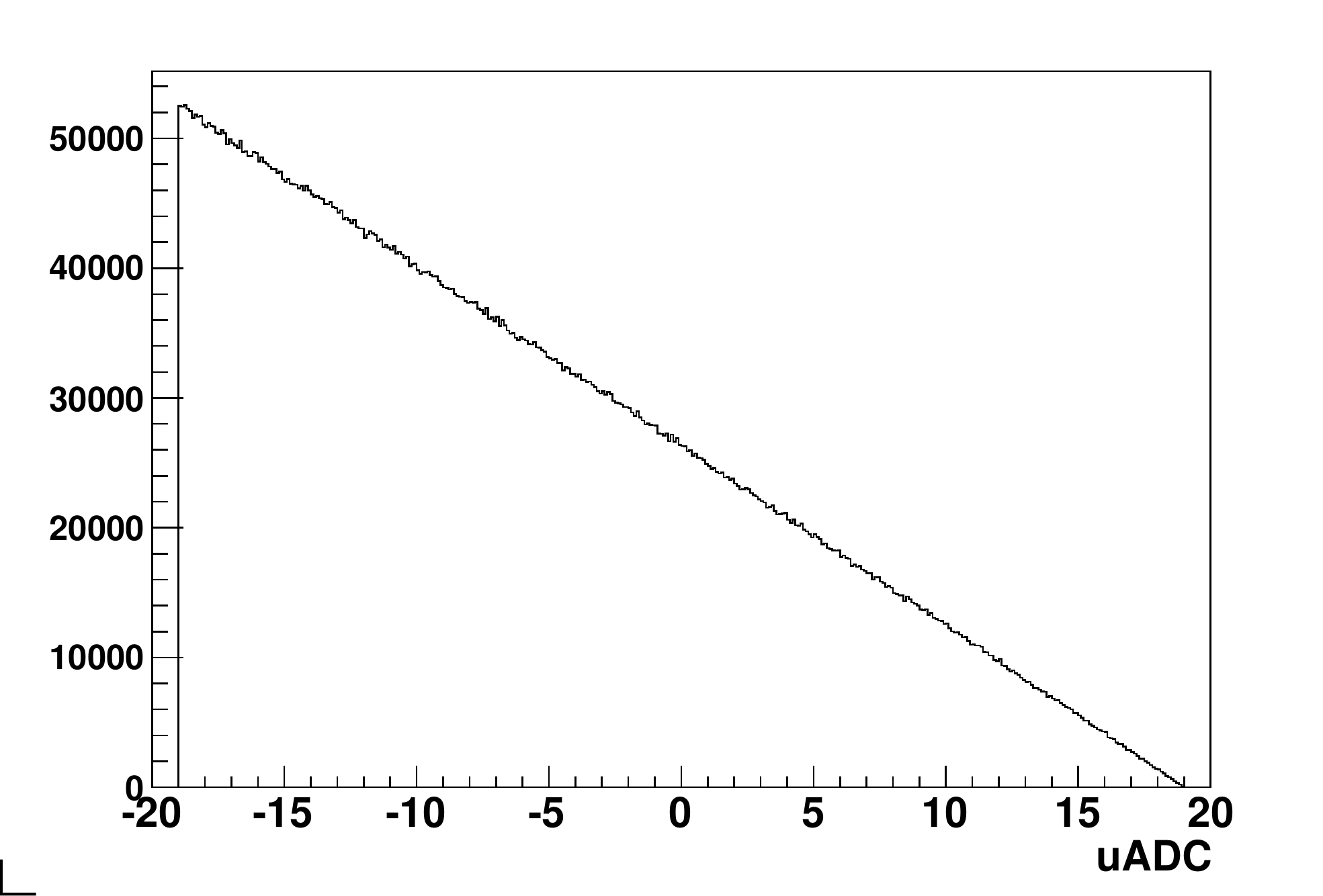}
 	\caption{Distribution of the non-Gaussian noise used in the simulation}
 	\label{fig:shape}}
 \end{figure}

In the reference configuration, the Gaussian distribution of C$^{1}$ is replaced by a non-Gaussian distribution. The non-Gaussian distribution is shown in figure~\ref{fig:shape}.

\begin{figure}[h]
 	{\centering 
 		\includegraphics[width=1.\textwidth]{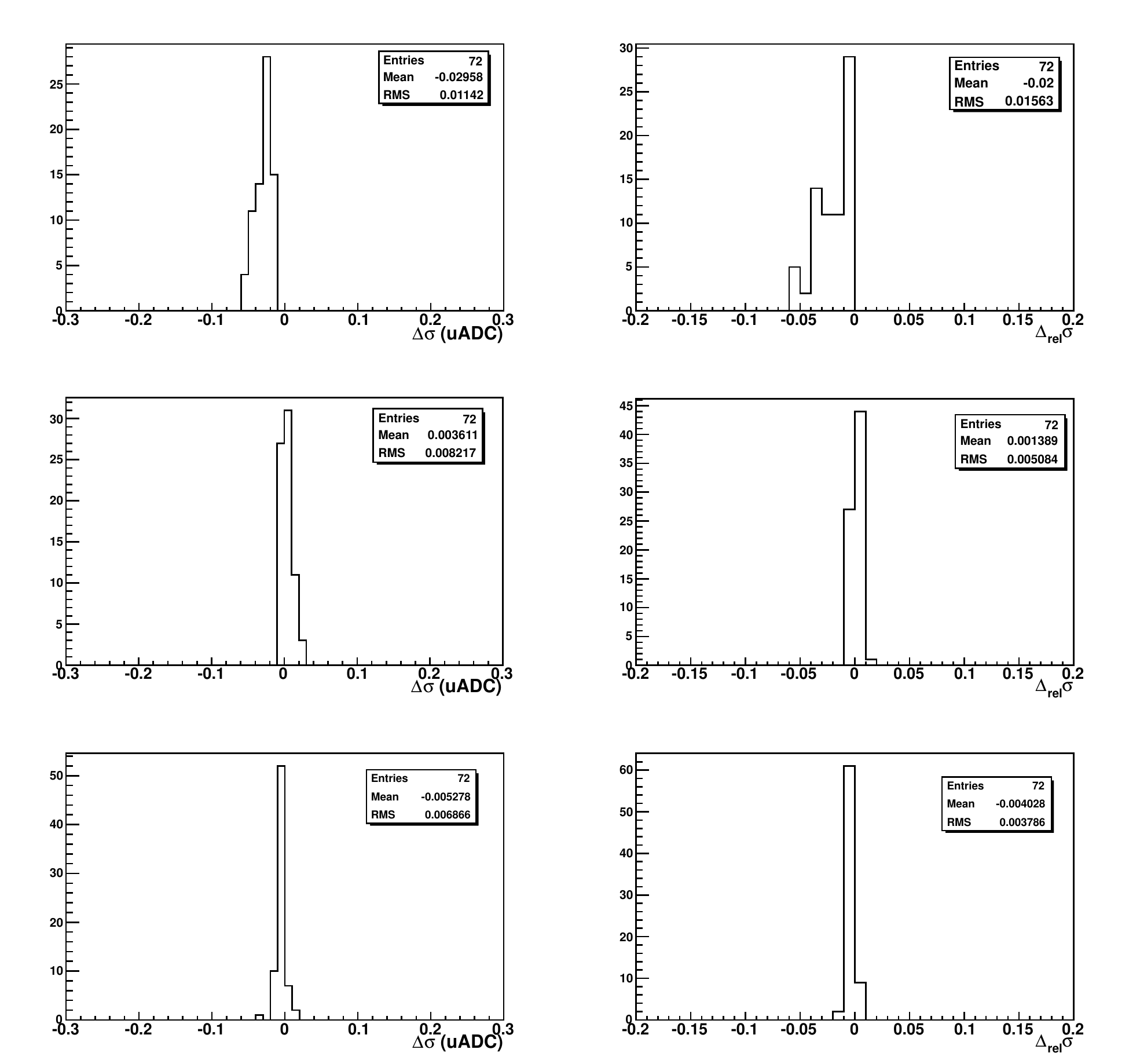}
 	\caption{Left side: $\Delta \sigma_{i}$ in ADC units - Right side: $\Delta_{rel} \sigma_{i}$ - Criteria for incoherent source, C$^{1}$ and C$^{2}$ are plotted respectively on the top, on the middle and on the bottom.}
 	\label{fig:resNG}}
 \end{figure}

Figure~\ref{fig:resNG} shows the robustness criteria for all channels. $\Delta \sigma_{i}$ is on the left and $\Delta_{rel} \sigma_{i}$ is on the right. Criteria for incoherent sources, C$^{1}$ and C$^{2}$ are plotted respectively on the top, on the middle and on the bottom.

The relative difference between simulated and calculated noise is always below than 5\% and $\left | \Delta \sigma\right | $ is below 0.1 uADC. The noise of $C^{1}$ is well reproduced. However, the noise distrubution shape cannot be extrapolated with this method.

\section{Conclusion}

We develop a method based on the study of the covariance matrix to identify and characterize Gaussian noise sources in multi-channel analysis. We apply this method to the SPIROC2 ASIC. The simulation study shows that the method is reliable. However the results may be improved using other minimization algorithm like simulated annealing to optimize the approximation of the system global optimum. Although this method has been developped for a specific application, it can be used for any multi channel analysis.

\bibliographystyle{unsrt_tibo}

\bibliography{biblio}

\end{document}